\documentclass [12pt,elsart,epsfig]{article}     
\usepackage{epsfig}
\begin{document}

\begin {center}

{\Large \bf A study of $\bar pp \to \eta \eta \eta$ for masses 1960 to 2410 
MeV/c$^2$}
\vskip 5mm
{A.V. Anisovich$^c$, C.A. Baker$^a$, C.J. Batty$^a$, D.V. Bugg$^b$,  
V.A. Nikonov$^c$, A.V. Sarantsev$^c$, V.V. Sarantsev$^c$, 
B.S.~Zou$^{b}$ \footnote{Now at IHEP, Beijing 100039, China} \\
{\normalsize $^a$ \it Queen Mary and Westfield College, London E1\,4NS, UK}\\
{\normalsize $^b$ \it Rutherford Appleton Laboratory, Chilton, Didcot OX11 0QX,UK}\\
{\normalsize $^c$ \it PNPI, Gatchina, St. Petersburg district, 188350, Russia}\\ 
[3mm]}
\end {center}

\begin{abstract}
Data on $\bar pp \to \eta \eta \eta$ for beam momenta 600--1940 MeV/c 
are presented.
The strongest channel is $f_0(1500)\eta$ from the initial $\bar pp$ 
state $^1S_0$.
Together with $\eta \pi ^0\pi ^0$ data, the $3\eta$ data determine the
branching ratio $BR[f_0(1500) \to \eta \eta]/BR[f_0(1500) \to \pi ^0\pi ^0]
= 0.42 \pm 0.09$.
They are consistent with a dominant contribution from an $I=0$, $C=+1$
$J^{PC} = 0^{-+}$ resonance observed earlier in the $\eta \pi ^0\pi ^0$ data;
from the combined $\eta \pi ^0\pi ^0$ and $\eta \eta \eta$ data, its mass is
$M = 2320 \pm 15$ MeV and its width $\Gamma = 230 \pm 35$ MeV.
\end{abstract}

As part of a study of $\bar pp$ annihilation in flight to neutral final states,
we have earlier presented results on $\bar pp \to \eta \pi ^0 \pi ^0 $ [1], 
with statistics of typically 70,000 events per momentum.
We have at the same time collected statistics of up to 192 events per
momentum on $\bar pp \to \eta \eta \eta$ in the $6\gamma$ channel.
Despite the limited statistics, some useful conclusions may be drawn.

In $\eta \pi ^0 \pi ^0 $ data, the $f_2(1270)\eta$ final state is dominant; 
the contribution from $f_0(1500)\pi$ is
small at all beam momenta: up to 3.2\% with 
errors of $\sim 0.6\%$ at each momentum.
However, the branching ratio of $f_0(1500)$ to $\eta\eta $ is a factor 
$\sim 45$ larger than for $f_2(1270)$; 
consequently $f_0(1500)$ makes the larger
contribution to $3\eta$ data.
These data therefore provide a valuable check on some features of 
the most recent amplitude analysis of $\bar pp \to \eta \pi ^0 \pi ^0$; 
this has recently been analysed in a combined fit with
data on $\bar pp \to \pi ^0 \pi ^0$, $\eta \eta$, $\eta \eta '$ and $\pi ^-
\pi ^+$ [2].
A $0^-$ resonance found at $\sim 2285$ MeV in that analysis is predicted to be
the main feature of the $f_0(1500)\eta$ channel here.
The present $3\eta$ data allow an improvement in the determination of its mass
and width.

The data were taken at LEAR using the Crystal Barrel detector.
Experimental details have been given in Ref. [1], and here we consider only
features which involve identification of the $3\eta$ final state.
In order to isolate this rare channel in $6\gamma$ data, tight selection
criteria are needed to eliminate backgrounds from other channels.
In processing data, we first demand exactly six photon showers, each
confined to a block of $3 \times 3$ adjacent CsI crystals of the detector.
Extra energy deposits produced by Compton scattering out of the primary
showers into nearby crystals (so-called 'split-offs') appear in $\sim 50\%$
of events.
All attempts to recover such events lead to backgrounds
higher by typically a factor 3-6.
We therefore reject events containing such additional
energy deposits.
Events where two photons from one  $\pi ^0$ merge into a single 
shower are also discarded.

A least squares kinematic fit is first required to $\bar pp \to 6\gamma$,
satisfying energy-momentum conservation with confidence level $CL >10\%$.
Then 7C fits are made to all 3-particle final states involving
$\pi ^0,~ \eta $ or $\eta '$.
We reject events  which fit $3\pi ^0$, $\eta \pi ^0 \pi ^0 $,
$\eta \eta \pi ^0$, $\eta '\pi ^0 \pi ^0 $ or $\eta \eta '\pi ^0$ with
$CL >10^{-4}$.
As further rejection against the prolific $3\pi ^0$,  $\eta \pi ^0 \pi ^0$
and $\eta \eta \pi ^0$ channels, 
events are rejected if they fit $\pi ^0 \pi ^0 \gamma \gamma$ or
$\pi ^0 \eta \gamma \gamma$ with  $CL >10^{-4}$.

\begin {table}[htp]
\begin {center}
\begin {tabular}{cc}
\hline
Channel & Background (\%)\\\hline
$\eta \eta \pi ^0$ & 0.1 \\
$\eta \pi ^0 \pi ^0 \pi ^0 $ & 0.1\\
$\eta \eta \pi ^0 \pi ^0$ & 1.1\\
$\omega \eta \eta, ~\omega \to \pi ^0 \gamma$ & 0.9\\
$\omega \eta \pi ^0, ~\omega \to \pi ^0 \gamma$ & 0.1\\\hline
\end {tabular}
\caption {Background levels in $6\gamma$ data from competing channels at
1800 MeV/c.}
\end {center}
\end {table}
A Monte Carlo study using 20,000 generated events for all competing channels
shows that the main sources of potential background are $\eta \eta \pi ^0$,
$\eta \pi ^0 \pi ^0 \pi ^0 $, $\eta \eta \pi ^0 \pi ^0 $, and
$\omega \eta \pi ^0$ or $\omega \eta \eta $ where
$\omega \to \pi ^0 \gamma$.
In the last four cases, two photons or one go undetected.
The background levels at 1800 MeV/c from competing channels are
estimated from this Monte Carlo simulation and are 
summarised in Table 1.
The principal backgrounds arise from
$\omega \eta \eta$ and $\eta \eta \pi ^0 \pi ^0$.
The situation is similar at other momenta and the total surviving
background is $2.3\%$ within errors at all momenta.
Kinematic fits to the $\eta \eta \gamma \gamma$ hypothesis are consistent with
this estimate and show only a few
scattered events with $M(\gamma \gamma )$ outside the 
$\eta \to \gamma \gamma$ peak. 

\begin {table}[htp]
\begin {center}
\begin {tabular}{cccc}
\hline
Momentum & $3\eta$  & $\epsilon $ & $\sigma (3\eta)$  \\
(MeV/c) & events & & $(\mu b)$ \\\hline
600 & 9 & 0.130 &  $ 3.0 \pm 0.7$ \\
900 & 68 & 0.126 &  $4.3 \pm 0.4$ \\
1050 & 72 & 0.126 &  $5.0 \pm 0.5 $ \\
1200 & 157 & 0.124 &  $6.9 \pm 0.5 $\\
1350 & 103 & 0.121 &  $6.4 \pm 0.5 $ \\
1525 & 99 & 0.117 &  $8.3 \pm 0.6 $ \\
1642 & 142 & 0.115 &   $9.8 \pm 0.6$ \\
1800 & 192 & 0.113 &   $9.5 \pm 0.5$ \\
1940 & 163 & 0.112 &   $5.7 \pm 0.4$ \\\hline
\end {tabular}
\caption {Numbers of $3\eta$ events after background subtraction and
reconstruction efficiency $\epsilon$;  the last column shows the 
weighted mean of the integrated cross section
for $3\eta$ (corrected for all $\eta$ decays) from $6\gamma$ and $10\gamma$
data. 
Errors include statistics and the uncertainty in normalisation of cross
sections at individual nomenta from Ref. [3]}
\end {center}
\end {table}

Columns 2 and 3 of Table 2 show
the number of surviving $3\eta$ events and
also the reconstruction efficiency $\epsilon$.
This efficiency is estimated by generating $\ge 50,000$ Monte Carlo events at
every momentum.
These events are subjected to identical selection procedures to data and
are used for the maximum likelihood fit described below.
Cross sections are derived from the number of observed events, corrected
by the efficiencies of Table 2 and using the number of incident antiprotons
and the target length of 4.4 cm.
A correction is applied for an observed dependence of reconstruction
efficiency as a function of beam rate; this is described in full in Ref. [3].

\begin{figure}
\begin{center}
\vskip -11mm
\epsfig{file=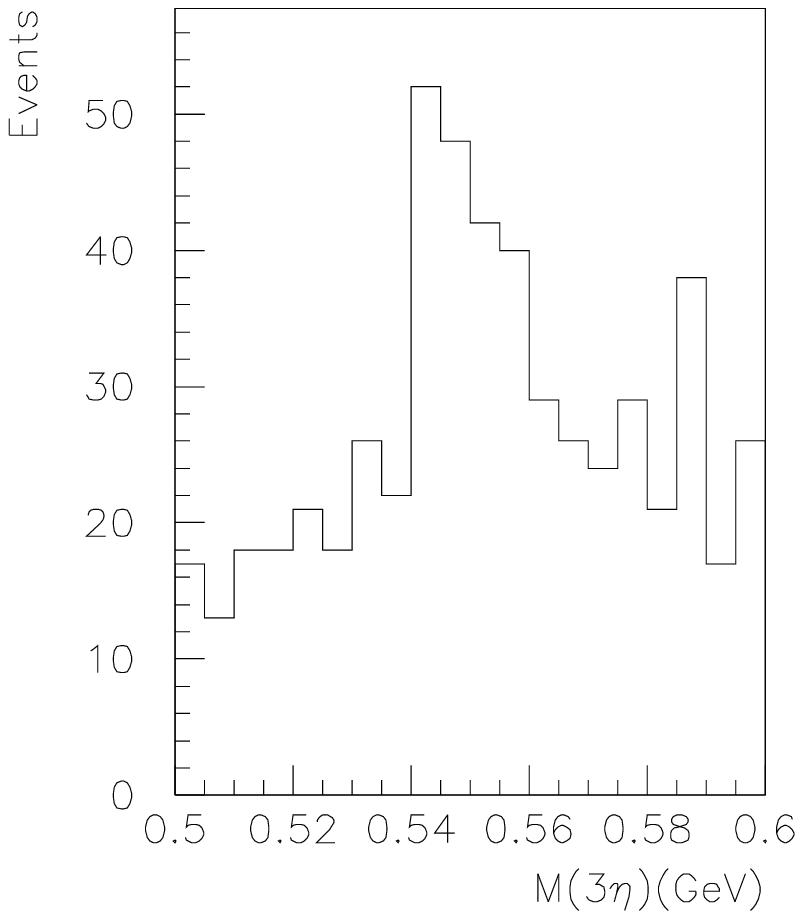,width=7.9cm}\
~\
\vskip -7.5mm
\caption{The $\eta \to 3\pi ^0$ peak in $\bar pp \to \eta \eta 3\pi ^0$ data at
900 MeV/c.}
\end{center}
\end{figure}
As a check, the $10\gamma$ data have also been examined for the final state
$\eta \eta 3\pi ^0$, where one $\eta \to 3\pi ^0$.
Events are selected by demanding exactly 10 photon showers  and a
kinematic fit to $\eta \eta 3\pi ^0$ with  $CL >10\%$.
Events fitting $5\pi ^0$ or $\eta 4\pi ^0$ with $CL >0.1\%$ are rejected.
Unfortunately, the background level under the $\eta \to 3\pi ^0$
peak is substantial (20\% of the signal at the highest momenta, rising
to 80\% at the lowest).
It arises mostly from $\eta \eta 3\pi ^0$ and cannot be reduced.
It is illustrated at 900 MeV/c in Fig. 1.
This background level is too high to allow a physics analysis.
However, these events may be used, after subtracting 
background,  to check the evaluation of integrated cross sections.
Fig. 2 compares cross sections for $\bar pp\to 3\eta$ 
from $6\gamma$ events (circles joined by the full curve) and
$10\gamma$ events (squares joined by the dashed curve).
There is agreement between these two determinations
within the errors.
The last column of Table 2 shows the weighted mean of the integrated cross sections 
from $6\gamma$ and $10\gamma$ data.
A detail is that statistics at 600 MeV/c are much lower than at other momenta;
this accounts for the small number of events, despite a similar cross section
to that at 900 MeV/c.
                                        
\begin{figure}
\begin{center}
\vskip -18mm
\epsfig{file=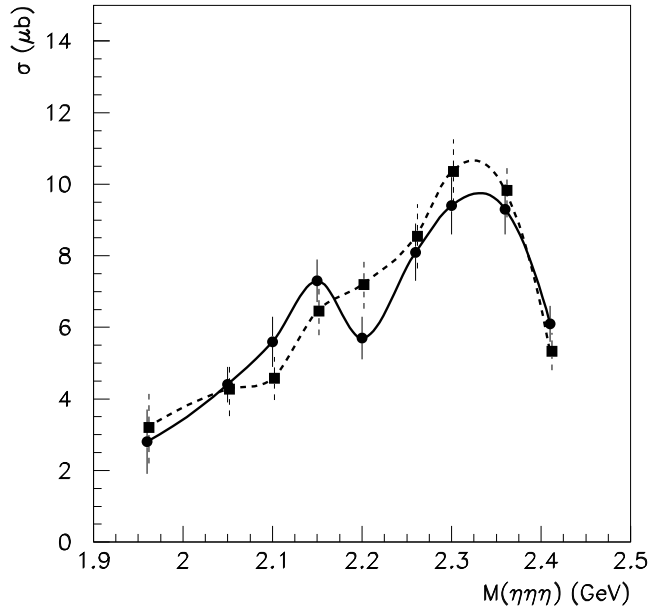,width=10cm}\
~\
\vskip -9mm
\caption {Comparison of integrated cross sections from $6\gamma$ data
(circles and full curve) with that from $10\gamma$ data (triangles and
dashed curve).}
\end{center}
\end{figure}

Dalitz plots from $6\gamma $ data and projections on to 
$M(\eta \eta )$ are shown at 8 beam momenta from 1940 to 900 MeV/c in 
Figs. 3 and 4.
Processes which ned to be considered are:
\begin {eqnarray}
\bar pp &\to& f_0(1500)\eta \\
        &\to& \sigma \eta \\
        &\to& f_2(1270) \eta \\
        &\to& f_0(1370) \eta \\
        &\to& f_2'(1525) \eta 
\end {eqnarray}
Of these, the last two turn out to be negligible.
The $f_0(1500)$ is conspicuous at the higher momenta.
The $f_2(1270)$ is obscured in mass projections of Figs. 3 and 4 
by reflections from the stronger $f_0(1500)$.
The fit shows that it makes a small but significant contribution.
The $\sigma \eta$ channel makes a strong contribution to 
$\eta \pi ^0 \pi ^0$ data, so its presence in $3\eta$ is necessary;
it accounts well  for the small uniform component over the Dalitz plots of
Figs. 3 and 4.
Here $\sigma$ is a shorthand for the $f_0(400-1200)$ of the Particle Data 
Group (PDG) [4]. 
It is fitted with the parametrisation of Zou and Bugg [5].

\begin{figure}
\begin{center}
\vskip -16mm
\epsfig{file=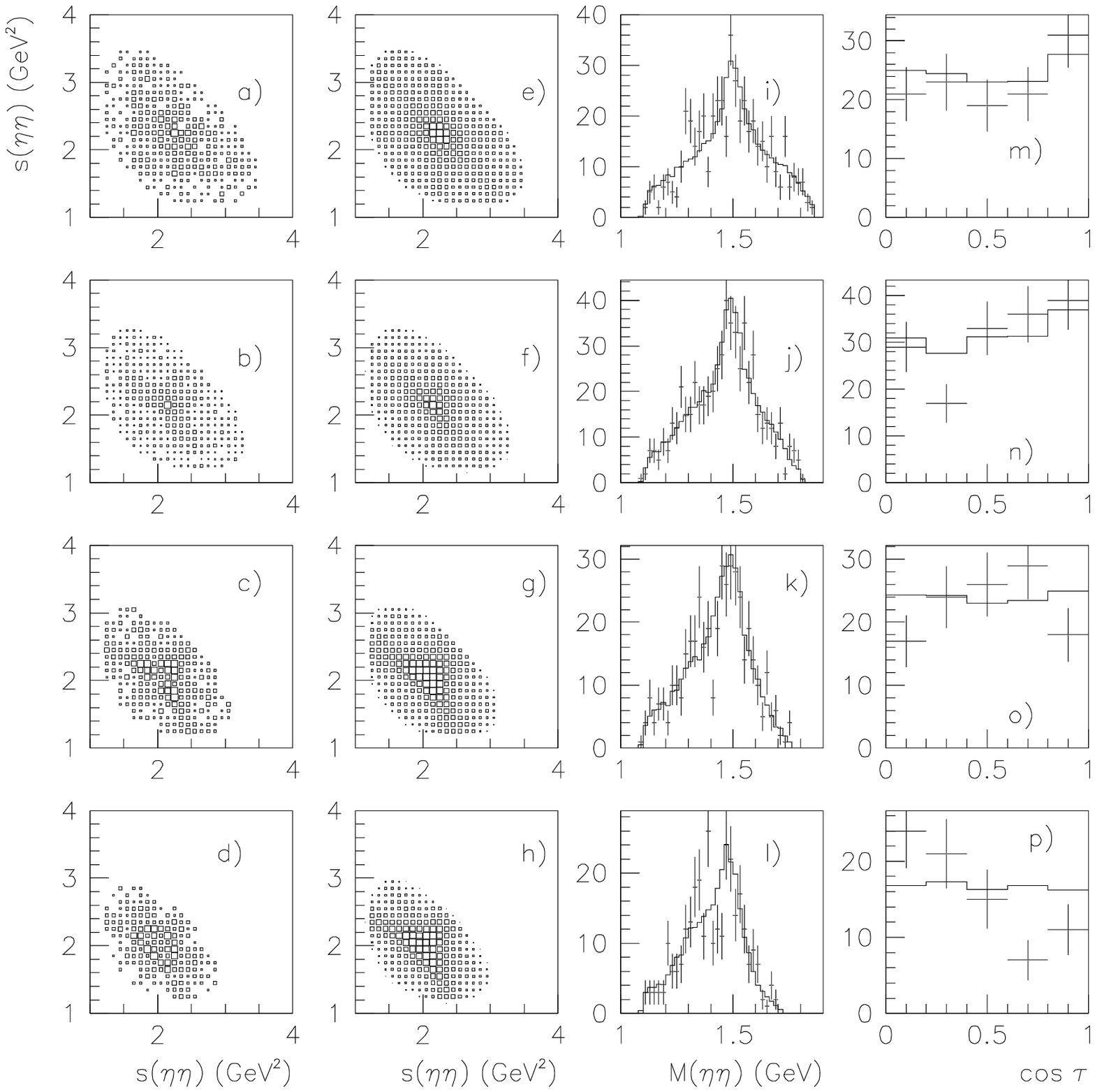,width=15.9cm}\~\
~\
\vskip -9mm
\caption {(a)--(d) Dalitz plots of data at beam momenta of
1940, 1800, 1642 and 1525 MeV/c,
(e)--(h)  fitted Dalitz plots, (i)-(l) projections on to
$M(\eta \eta )$, (m)--(p) production angular distribution for events
in a band 120 MeV wide centred on $f_0(1500)$. Histograms show the fit.}
\end{center}
\end{figure}
\begin{figure}
\begin{center}
\vskip -16mm
\epsfig{file=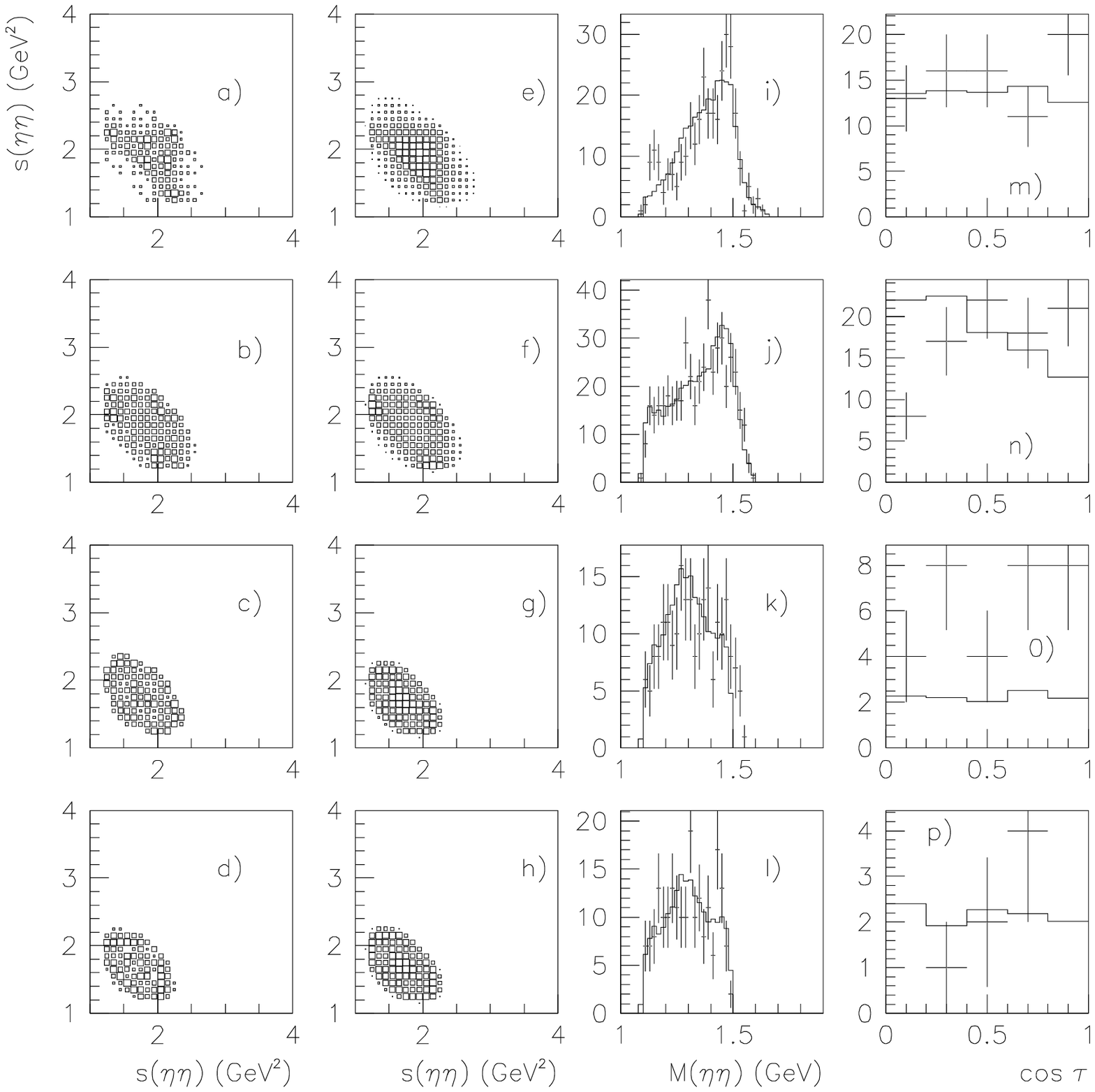,width=15.9cm}\
~\
\vskip -9mm
\caption{As Fig. 1 at 1350, 1200, 1050 and 900 MeV/c.}
\end{center}
\end{figure}

The data have been fitted by the maximum likelihood method.
Present data are closely related to those for $\bar pp \to \eta \pi ^0\pi ^0$,
where statistics are much higher.
Our strategy is to take magnitudes and phases of all
partial waves for $f_2(1270)\eta$, $f_0(1500)\eta$ and $\sigma \eta$
from our analysis of
$\eta \pi ^0 \pi ^0 $ [2], allowing one free parameter to determine the
branching ratio 
$r_{1270} = BR[f_2(1270) \to \eta \eta]/BR[f_2(1270) \to \pi ^0 \pi ^0 ] $, 
a second to determine the  branching ratio 
$r_{1500} = BR[f_0(1500) \to \eta \eta]/BR[f_0(1500) \to \pi ^0 \pi ^0 ] $
and a third to determine the branching ratio 
$r_{\sigma } = BR[\sigma \to \eta \eta]/BR[\sigma \to \pi ^0 \pi ^0 ] $.
The contribution from $f_2(1270)$ is small, as shown below in Fig. 6(b)
by the chain curve.
The value of $r_{1270}$ can therefore be set to the PDG value.
The absolute value of the integrated cross section for $3\eta$ is sensitive
to $r_{1500}$ and $r_\sigma$, where therefore need to be fitted.

\begin {table}[htp]
\begin {center}
\begin {tabular}{ccccc}
\hline
Momentum & $-f_0(1500)\eta$ & $-f_2(1270)\eta$ & $+f_0(1370) \eta $ &
$+f_2'(1525)\eta$ \\\hline
900  &-27.6  &-4.4  & 1.1 & 1.2 \\
1050 &-25.9  &-10.8 &1.8  & 0.9\\
1200 &-44.9  &-1.0  &1.0  & 0.1\\
1350 &-56.6  &-7.1  &2.2  & 0.8\\
1525 &-15.0  &-19.8 &0.8  & 1.7\\
1642 &-21.8  &-41.6 &4.4  & 2.0\\
1800 &-95.2  &-20.6 &0.3  & 0.5\\
1942 &-50.6  &-11.5 &4.8  & 1.1\\\hline
\end {tabular}
\caption {Changes in log likelihood (a) removing $f_0(1500)\eta$
from the fit, (b) removing $f_2(1270)\eta$, (c) adding $f_0(1370)\eta$ and
(d) adding $f_2'(1525)\eta$.}
\end {center}
\end {table}
Columns 2 and 3 of Table 3 show changes in log likelihood when $f_0(1500)$ or
$f_2(1270)$ is removed from the fit at individual momenta. 
Our definition is such that a change of
log likelihood of 0.5 corresponds to a one standard deviation change for one
degree of freedom, so the observed changes indicate the presence
of both channels.
Dropping $f_2(1270)$ from the overall fit to all momenta, log likelihood is
worse by 90.4.

Histograms on Figs. 3 and 4 display results of the fit to all momenta
simultaneously. The first column
of Dalitz plots shows data and the second column the fits.
Every event is plotted in three $\eta \eta$ combinations.
The low statistics cause substantial fluctuations for data and are responsible
for some apparent disagreement with the fit; the latter is much smoother 
than data since Monte Carlo statistics of over 5000 events per momentum 
are used in the maximum likelihood fit.
In fact, the $\chi ^2$ between data and fit on the Dalitz plot averages 1.1 per
point, so there is no discrepancy.

The partial wave amplitude for production of the $f_2$ 
between particles 1 and 2 will be used as an illustration of the
way amplitudes are parametrised. It takes the form
\begin {equation}
f = \frac {G}{M^2 - s_{12} - iM\Gamma} B_2(k)B_L(p)Z(p,k).
\end {equation}
Here $G$ is a complex coupling constant, 
the denominator refers to the $f_2(1270)$, $s$ is   mass squared for
the $\eta _1\eta _2$ combination,  and $Z$ is a relativistic 
Zemach tensor given
explicitly in Ref. [1]. The cross section is obtained from the coherent sum of
amplitudes for the three $\eta \eta$ combinations.
The value of $G$ for $3\eta$ data is related to that for $\eta \pi ^0 \pi ^0$
by $G(3\eta ) = \sqrt {r_{1270}k_{\pi }/k_{\eta }}
G(\eta \pi ^0 \pi ^0)/\sqrt {3}$.
The factor $1/\sqrt {3}$ allows for the three identical $\eta \eta$ pairs;
the ratio $\sqrt {k_{\pi }/k_{\eta }}$, involving momenta $k$ of $\pi$ and
$\eta$ in the decay $f_2 \to \pi \pi$ and $\eta \eta$, allows for phase space
for those decays.
Standard Blatt-Weisskopf centrifugal barrier factors $B_L$ with radius 0.8 fm
are used to parametrise the dependence on the centre of mass momentum $p$
with which the resonance is produced, with orbital angular momentum $L$, 
and also on its decay momentum $k$ in the resonance rest frame.

The complex coupling constant $G$ assigns a phase to each partial wave.
This phase originates from the initial state $\bar pp$ interaction and from
rescattering in the final state.
The approximation we adopt is that the phase is the same for decays $f_2(1270)
\to \eta \eta$ and $f_2(1270) \to \pi ^0 \pi ^0$
and likewise for $f_0(1500) \to \eta \eta$ and $\pi ^0 \pi ^0$.
This is in the spirit of the isobar model, where it is assumed that the final
resonant state is reached after any number of intermediate rescatterings
and then decays into the $\eta \eta$ and $\pi ^0 \pi ^0 $ channels with
the same phase. 
Relative magnitudes of $f_2(1270)$ contributions in different partial waves are
taken from the fit to $\eta \pi ^0 \pi ^0$ data of Ref. [2];
likewise for $f_0(1500)$ and $\sigma$ contributions. 
However, rescattering is
possible in $\eta \pi ^0 \pi ^0 $ to additional final states
$a_2(1320)\pi ^0$ and $a_0(980)\pi ^0$; therefore we allow an overall phase 
difference of $f_2(1270)\eta$ relative to $f_0(1500)\eta$ and $\sigma \eta$ 
at all momenta between  $\eta \pi ^0 \pi ^0$ data and $3\eta$.
The fit to $3\eta$ is therefore related to that to $\eta \pi ^0\pi ^0$ by 4
parameters: two relative phases between $f_2(1270)\eta$, $f_0(1500)\eta$
and $\sigma \eta$ and two branching ratios $r_{1500}$ and $r_\sigma$, common
to all momenta. 
Because the contribution from $f_2(1270)$ is small, results concerning
$f_0(1500)\eta$ have little sensitivity to the detailed partial wave
decomposition of $f_2(1270)\eta$. 
This is because correlations between
the two channels arise only where bands cross on the Dalitz plots and
involve rather few events.

The prediction from $\eta \pi ^0 \pi ^0$ data [2] is tha the dominant
contribution to $3\eta$ will be from $L=0$ decays of a $0^-$ resonance at
2285 MeV. 
A small $L=2$ component is predicted, roughly $15\%$ of the intensity of
$L=0$. 
Contributions with $L=1$ and 3 are found in Ref. [2] to be very small.
For $f_0(1500)\eta $ final states, the angular distributions against centre of
mass production angle $\tau$ are shown in Figs. 3 and 4. 
Fortuitously, the $L=0$ and $L=2$ amplitudes are
nearly orthogonal, with the result that interferences between them are small. 
Hence angular distributions in Fig. 3 and 4 are close to isotropic. 
The discrepancy between data and fit in Figs. 4(o) is associated with the 
very small number of  events in the $f_0(1500)$ mass region.

A point with interesting consequences is that the three $f_0(1500)$ bands
cross at the centre of the Dalitz plot near a beam momentum of 1800 MeV/c.
The strong constructive interference observed in Fig. 3(b) between the three bands
requires that the process goes largely through a single $\bar pp$ partial
wave with even $L$.
Amplitudes for production with $L = 1$ cancel at the intersection point of
the three bands and their sum changes sign about this point.
To see this, consider $f_0(1500)$ production from the initial state $^3P_1$.
For initial helicity $m = 0$, the Clebsch-Gordan coefficient for coupling
to $\bar pp$ is zero.
For $m = 1$, the amplitude is proportional to $p \sin \tau \exp (i\phi )$,
where $\tau$ is the polar angle for production of the resonance and
$\phi$ is the associated azimuthal angle.
This amplitude is therefore proportional to $p_X + ip_Y$, where $p_{X,Y}$ are
transverse momentum components of the resonance.
For the coherent superposition of the three $f_0(1500)\eta$ combinations,
the resultant at the intersection of the three bands is zero, by momentum
conservation.
The same result extends to $L = 3$ for $f_0(1500)\eta$ and to
production of $f_2'(1525)$ with $L = 1$ and 3,
though the amplitude then contains additional factors for the $f_2'$ decay.
Fig. 5 shows the predicted Dalitz plot for $^3P_1 \to f_0(1500)\eta$ at
1800 MeV/c.
This distinctive pattern is absent from the data of Fig. 3.
The amplitude analysis confirms that any $L = 1$ or 3 processes producing
$f_0(1500)$ or $f_2'(1525)$ at 1642, 1800 and 1940 MeV/c are absent or
very weak (summed cross sections $\leq 10\%$ of $L = 0$).
\begin{figure}
\begin{center}
\vskip -16mm
\epsfig{file=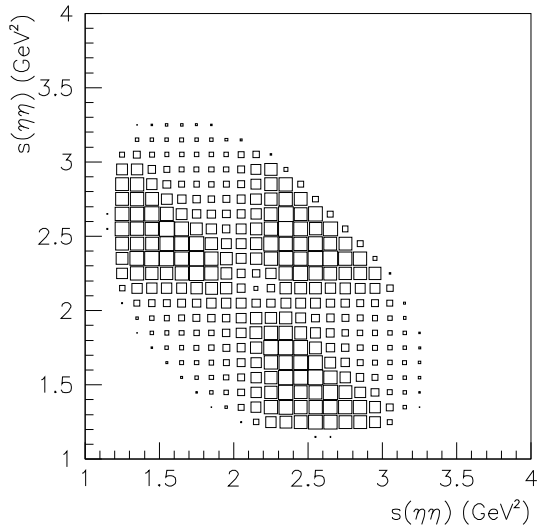,width=8.9cm}\
~\
\vskip -9mm
\caption{The predicted Dalitz plot at 1800 MeV/c for $^3P_1 \to f_0(1500)\eta$.}
\end{center}
\end{figure}

The essential physics conclusions of the amplitude analysis 
are illustrated in Fig. 6.
Fig. 6(a) shows a fit (with details given below) to the 
integrated $3\eta$ cross section averaged between $6\gamma$ and $10\gamma$ data.
The peak requires, as predicted,  a dominant contribution from a $J^P = 0^-$ resonance
in the $f_0(1500)\eta$ channel; its intensity  is shown
by the full curve in Fig. 6(b).
It appears at a slightly higher mass than predicted: $2328 \pm 16$ MeV compared
with $M = 2285 \pm 20$ MeV of Ref. [2].
The shift is compatible with the combined errors of the two analyses.
The width is discussed below.

In addition, a $\sigma \eta$ contribution is required, as shown by the
dashed curve.
This interferes destructively with $[f_0(1500)\eta ]_{L = 0}$.
Thirdly, there is a small contribution, shown dotted, from $\eta _2(2267) \to
[f_0(1500)\eta ]_{L = 2}$.
The $f_2(1270)\eta$ channel makes a small contribution 
$< 1.3 \mu b$ for all masses (chain curve).

The $f_2\eta$ contribution contains too many partial waves and is too small
to allow a useful determination of the branching ratio $r_{1270}$ between
$\eta \eta$ and $\pi ^0 \pi ^0$.
This parameter is therefore fixed at the PDG value, $1.35 \times 10^{-2}$ [4].
The corresponding ratio $r_{1500}$ for $f_0(1500)$ may be fitted over a range
of values, because of interference of this channel with $\sigma \eta$.
The width of the $0^-$ resonance at $2330$ MeV correlates strongly
with $r_{1500}$.
Small $r$ values require a narrow width for the resonance, so as to reproduce
the peak cross section. 
Amsler et al. [6] determine from $\bar pp \to \eta \eta \pi ^0$ and $3\pi ^0$
data at rest a ratio $r_{1500} = 0.47 \pm 0.21$;
using this value, the width of the $0^-$ resonance optimises at $\Gamma = 240$
MeV, distinctly smaller than the value $325 \pm 30$ MeV found from $\eta \pi ^0
\pi ^0$ data [2].
The WA102 collaboration [9] finds $r_{1500} = 0.54 \pm 0.09$, compatible with
Ref. [6] and with our determination below.
Abele et al. [6] find $r_{1500} = 0.23 \pm 0.04$, which requires $\Gamma =
154$ MeV for the $0^-$ resonance. 
This narrow width gives a significantly
poorer log likelihood in fitting $3\eta$ data and is hard to reconcile 
with $\eta \pi ^0 \pi ^0$ data.
\begin{figure}
\begin{center}
\vskip -18mm
\epsfig{file=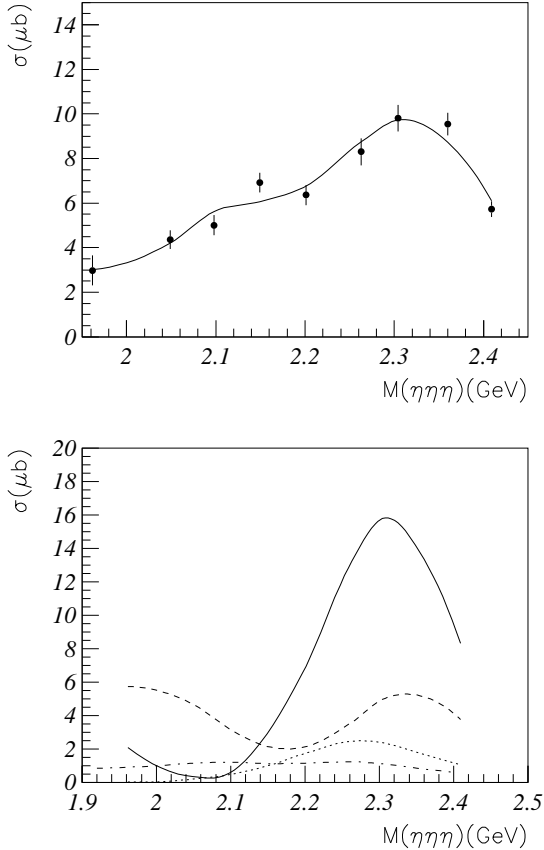,width=8.9cm}\
~\
\vskip -9mm
\caption{(a) The fit to the mean integrated cross section from $6\gamma$
and $10\gamma$ data; (b) the contributions from
$\eta (2320) \to  [f_0(1500)\eta]_{L = 0}$ (full curve), 
$\sigma \eta$ (dashed),
$\eta_2 (2267) \to  [f_0(1500)\eta]_{L = 2}$ (dotted), and
$f_2(1270)\eta$ (chain curve).} 
\end{center}
\end{figure}

   We find that the fit to $\eta \pi ^0 \pi ^0$ data will adjust towards these
new values of $M$ and $\Gamma$ with only a minor change in log likelihood and
in other partial waves.
The optimum fit to combined $3\eta$ and $\eta \pi ^0 \pi ^0$ data gives $M =
2320 \pm 15$ MeV, $\Gamma = 230 \pm 35$ MeV and 
\begin {equation}
r_{1500} = 0.39 \pm 0.09.
\end {equation}
These values are used in the curves of Fig. 6. 
Intensities derived from each component are shown in Fig. 6(b) 
including interferences between all three $\eta \eta$ channels and 
after integrating over the Dalitz plot.
In comparing with corresponding intensities for the $\eta \pi \pi$ channel,
shown in Fig. 2 of Ref. [2], one must take note of two points.
Firstly, strong constructive interferences between the three $\eta \eta$
channels in $3\eta$ enhance the peaks of Fig. 6(b), for example by 60\% for
$[f_0(1500)\eta ]_{L = 0}$.
Secondly, Fig. 6(b) uses a width of 230 MeV for the resonance, while Fig. 2 of
Ref. [3] uses 325 MeV; this increases the peak height in Fig. 6(b) by a
further factor 1.9.

   The $\sigma \eta$ intensity in Fig. 6(b) is $\sim 20\%$  of that for $\eta
\pi ^0 \pi ^0$.
If the $\sigma$ is dominantly non-strange, one expects from the composition of
the $\eta$, namely 
$$|\eta > = 0.8|(u\bar u + d\bar d)/\sqrt {2}> - 0.6|s\bar s>,$$
a ratio between $\eta \eta$ and $\pi ^0 \pi ^0$ decay of 
$(0.8)^4<k_{\eta }/k_{\pi }>$; 
here $k_{\eta ,\pi }$ are the momenta in the $\sigma$ rest frame in
$\eta \eta$ and $\pi \pi$ decays respectively.
Averaging over phase space for the $\sigma$, one finds 
$<k_{\eta }/k_{\pi }> \simeq 0.6$, and hence a predicted intensity for $\sigma
\eta$ in $3\eta$ of 25\% of that in $\eta \pi ^0 \pi ^0$, close to the fit.

From present data, we are unable to demonstrate  the presence
of the $J^P = 2^-$ resonance decaying to $[f_0(1500)\eta ]_{L = 2}$.
Deviations from isotropy in angular distributions of Figs. 3 and 4 are too small
to demonstate its presence, though angular distributions are compatible
with prediction within the large errors. 
We remark that the $2^-$ resonance is observed clearly in data on 
$\bar pp \to \eta '\pi ^0 \pi ^0$, 
where it makes a dominant contribution in  
decays to $f_2(1270)\eta '$ [9]. 

The fourth and fifth columns of Table 3 show changes in log likelihood when either
$f_0(1370)\eta$ or $f_2'(1525)$ is added to the fit with $L=0$.
Statistically one expects an improvement of 1 for two extra fitted parameters.
The evidence for either $f_0(1370)$ or $f_2'(1525)$ being present is insignificant.
Let $\alpha$ be the decay angle with respect to the beam in the rest fram of the
resonance.
The decay angular distribution $(3\cos ^2 \alpha -1)$ for $f_2'(1525)$ produced
from the initial state $^1D_2$ with $L=0$ is distrintive and the amplitude would
interfere with $f_0(1500)\eta$.
The $L=1$ production amplitude likewise gives a distinctive contribution to the
Dalitz plot, similar to that of Fig. 5.
The mean fitted value of its cross section is $15\%$ of $f_0(1500)\eta$ but some
or all of this is undoubtedly statistical noise in the fit.

From earlier analysis of $\bar pp \to \eta \eta \pi ^0$ and $3\pi ^0$
at rest, the ratio $BR[f_0(1370) \to \eta \eta ]/
BR[f_0(1370) \to \pi ^0 \pi ^0 ] = 0.056 \pm 0.04$ [6].
Despite the sizeable error, there was no doubt of the presence of some
$f_0(1370) \to \eta \eta$ signal in the $\eta \eta \pi ^0$ data [7], 
considerably higher than that for $f_2(1270)$.
From the small $f_0(1370)\eta$ signal in $3\eta$, we therefore deduce that
the signal observed in $\eta \pi ^0 \pi ^0 $ must be almost entirely
$f_2(1270)\eta$ with very little $f_0(1370)\eta$.
This is a useful check on the analysis of $\eta \pi ^0 \pi ^0$, where a small
$f_0(1370)$ signal is hard to identify in the presence of a dominant
$f_2(1270)\eta$ contribution.

In summary, despite low statistics, four results emerge from this analysis.
\begin {itemize}
\item The $3\eta$ and $\eta \pi ^0 \pi ^0$ data together require an $ I = 0$,
$C = +1$, $J^{PC} = 0^{-+}$ resonance with $M = 2320 \pm 10$ MeV, 
$\Gamma = 230 \pm 35$ MeV, decaying to $f_0(1500)\eta$.
\item The ratio of intensities in $3\eta $ and $\eta \pi ^0\pi ^0$ gives
$r_{1500} = 0.39 \pm 0.09$.
\item The ratio of $\sigma \eta$ intensities in $3\eta$ and $\eta \pi ^0 \pi
^0$ is close to that expected for non-strange composition of $\sigma \equiv
f_0(400-1200)$.
\item The absence of $f_0(1370) \to \eta \eta$ in $3\eta$ data is a useful
check that it makes very little contribution to $\eta \pi ^0 \pi ^0$ data.
\end {itemize}

We wish to thank the technical staff of the LEAR machine group
and of all the participating institutions for their invaluable
contributions to the success of the experiment. We thank the Crystal Barrel 
group for allowing use of the data. 
We acknowledge financial support from the
British Particle Physics and
Astronomy Research Council (PPARC).
The St. Petersburg group wishes to acknowledge financial support from PPARC and
INTAS grant RFBR 95-0267.


\begin {thebibliography}{99}
\bibitem {1} A.V. Anisovich  et al, Phys. Lett. B452 (1999) 173;
Nucl. Phys A651 (1999) 253.
\bibitem {2} A.V. Anisovich et al., {\it $I = 0$ $C = +1$ mesons from
1920 to 2410 MeV/c$^2$}, Phys. Lett. B (in press).
\bibitem {3} A.V. Anisovich et al., Phys. Lett. B468 (1999) 304;
Nucl. Phys. A662 (2000) 344.
\bibitem {4} Particle Data Group, Euro. Phys. J. C3 (2000) 1.
\bibitem {5} B.S. Zou, D.V. Bugg, Phys. Rev. D48 (1993) R3948.
\bibitem {6} C. Amsler et al., Phys. Lett. B355 (1995) 425.
\bibitem {7} D. Barberis et al., Phys. Lett. B479 (2000) 59.
\bibitem {8} A. Abele et al., Nucl. Phys. A609 (1996) 562.
\bibitem {9} A.V. Anisovich et al., {\it Data on $\bar pp \to 
\eta '\pi ^0 \pi ^0$ for masses 1960 to 2410 MeV/c$^2$}, 
Phys. Lett. B (in press).
\bibitem {10} C. Amsler et al., Phys. Lett. B353 (1995) 571.


\end {thebibliography}
\end{document}